\documentclass[fleqn,twoside]{article}
\usepackage{amsmath,amssymb}
\usepackage{espcrc2}

\usepackage{graphicx}
\usepackage{color}
\definecolor{gray}{rgb}{0.6,0.6,0.6}
\usepackage{feynmp}

\newcommand{\psibar}{{\overline{\psi}}}

\title{Taste-Changing in Staggered Quarks}
\author{Quentin Mason\address[Cornell]{Laboratory of Elementary-Particle Physics, Cornell University, Ithaca, NY 14853, USA}, Peter Lepage\addressmark[Cornell], Paul Mackenzie\address{Theoretical Physics Group, Fermilab P.O. Box 500, Batavia, IL 60510, USA}, Howard Trottier\address{Physics Department, Simon Fraser University, Burnaby, B.C.  V5A 1S6, Canada}, Joachim Hein\address{School of Physics, The University of Edinburgh, Edinburgh EH9 3JZ, UK}, Christine Davies\address[Glasgow]{Department of Physics \& Astronomy, The University of Glasgow, Glasgow G12 8QQ, UK}, and Eduardo Follana\addressmark[Glasgow], HPQCD collaboration.}

\begin{document}
\begin{abstract}
We present results from a systematic perturbative investigation of taste-changing in improved staggered quarks.  We show one-loop taste-changing interactions can be removed perturbatively by an effective four-quark term and calculate the necessary coefficients.
\vspace{1pc}
\end{abstract}
\maketitle
\section{INTRODUCTION}
\subsection{Motivation}
Accurate lattice simulations of the Standard Model require three light flavours of dynamical quarks.  The Improved Staggered-quark formalism is the only one capable of delivering large numbers of configurations with small quark masses anytime in the near future.  This formalism, however has many non-degenerate pions whose masses do \emph{not} vanish for zero quark mass.  The residual masses come from mixing between the staggered copies of the quarks, and vanish like $a^2$.

Na\"\i ve staggered quarks suffer from poorly convergent perturbative expressions and large pion splittings which can be suppressed by the use of fat links~\cite{Blum:1996uf}.  Staggered quarks which are improved to $\mathcal{O}(\alpha_s a^2,a^4)$~\cite{Lepage:1998vj} significantly reduce splittings in the pion spectrum~\cite{Toussaint:2001zc} and have small renormalisations~\cite{Hein:2001kw}.  A related scheme with small pion splittings is the HYP action~\cite{Anna}.

To further reduce these undesirable splittings there are three alternatives: (I) to reduce $a$ at considerable cost; (II) to correct for them using modified chiral perturbation theory~\cite{Lee:1999zx,Bernard:2001yj} or (III) to further improve the quark action.  In this work we show how to quantify and reduce the effects of these interactions by further improvement.  Firstly we look at taste-changing in na\"\i ve quarks where one has physical intuition and the calculation is easier, then we convert to staggered quarks and present our results and conclusions.

\section{TASTE-CHANGING}
\subsection{Na\"\i ve Quarks}
\begin{figure}[t]%
\begin{center}%
    \begin{fmfgraph*}(60,60)%
      \fmfset{arrow_len}{2mm}%
    \fmftopn{l}{2}%
    \fmfbottomn{r}{2}%
    \fmflabel{\small $0$}{l1}%
    \fmflabel{\small $\zeta\pi/a$}{l2}%
    \fmflabel{\small $0$}{r1}%
    \fmflabel{\small $\zeta\pi/a$}{r2}%
    \fmf{fermion,f=(0.45,,0.45,,0.45)}{l1,v1}%
    \fmf{fermion}{v1,l2}%
    \fmf{fermion}{r2,v2}%
    \fmf{fermion,f=(0.45,,0.45,,0.45)}{v2,r1}%
    \fmf{boson,f=(0.7,,0.7,,0.7),label=\small $\zeta\pi/a$}{v1,v2}%
    \fmfdot{v1,v2}%
    \end{fmfgraph*}%
\vspace{-2mm}
\caption{\label{F:treetastechange}Generic tree-level taste-changing diagram for massless na\"\i ve quarks}%
\end{center}%
\end{figure}
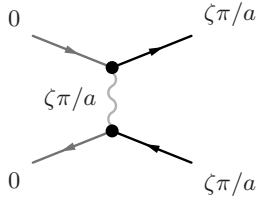

Na\"\i ve quarks have an exact symmetry:
\begin{equation}
  \text{quark}(p\sim 0)\quad\equiv\quad \text{quark}(p\sim\zeta\pi/a)
\end{equation}
where $\zeta=(1,0,0,0)$, $(1,1,0,0)$, \ldots\; giving 16 degenerate copies of every quark, one in each corner of the Brillouin zone.  These doublers can mix by exchanging a very hard gluon as shown in figure~\ref{F:treetastechange}.  This gluon is highly virtual with momentum $\mathcal{O}(\pi/a)$ and thus the quark-quark interaction is effectively a purely perturbative contact interaction at typical lattice spacings.  
\begin{fmffile}{boston_writeup_feyn}%

The tree-level interaction of figure~\ref{F:treetastechange} was understood and completely removed with the improved staggered action by using smearing to suppress high-momentum gluon emission from quarks~\cite{Blum:1996uf,Lepage:1998vj}.  To further reduce the splitting we turn to one-loop taste-changing diagrams of which there are five for massless quarks.  Each quark line has an odd number of gamma-matrices and therefore the spinor structure must be a combination of $\gamma_\mu, \gamma_{5\mu}$; while the colour structure is singlet or octet.  In principle there is a different coefficient for every current contact interaction between all combinations of the four corners of the Brillouin zone, 16$^4$ in total, although momentum conservation eliminates some.  The doubling transformation can be applied separately to each quark line, however, and along with rotational invariance leaves only four independent interactions labelled by $\zeta^2$ in each of the two spinor and two colour channels.  

\subsection{One-loop Corrections}
The calculations are easiest using (improved) na\"\i ve quarks, and na\"\i ve quark operators are easily converted to the familiar \emph{spinor}$\otimes$\emph{taste} staggered operators.  Ignoring gluons we have charge eigenstate currents:
\begin{align}
&\left.\psibar\; \gamma_n\,\otimes\,\xi_s\;\psi\right|_{\text{staggered}} \equiv\notag\\
&(-)^{\overline{s}\cdot x}\frac\eta2\left[\psibar(x)\gamma_s^\dagger\gamma_n\psi(x+\Delta x_{ns})\pm \text{h.c.}\right],
\end{align}
where $n, s$ are 4-vectors of 0's and 1's (like $\zeta$), $\overline{s} = \sum_{\nu\neq\mu}s_\nu\pmod 2$,
\begin{align}
\gamma_n = \prod_{\mu=0}^{3}\left(\gamma_\mu\right)^{n_\mu},
\end{align}
with $\Delta x_{ns}=n+s\pmod 2$ and $\eta$ a phase factor for hermiticity.  The $n\!\otimes\!s$ current corresponds to na\"\i ve momentum transfer $\zeta\pi/a=\overline{s}\pi/a$.  
Armed with our map we can translate all the na\"\i ve interactions to get the most general counter-terms:
\begin{align}
\Delta\mathcal{L} &=\alpha_s^2 \sum_{s,t}\frac12(\text{coeff})^2 \Bigl[\psibar\, \gamma_s\otimes\xi_t\,\psi\Bigr]^2 \notag\\
&+\alpha_s^2 \sum_{s,t,a}\frac12(\text{coeff})^2 \Bigl[\psibar\,\gamma_s\otimes\xi_t\,T^a\,\psi\Bigr]^2,
\end{align}
where the sum is over all one- and three-link currents which are Hermitian taste non-singlets with colour singlets and octets.  Adding these terms to the tree-level action will cancel the one-loop taste-changing and suppress the pion splittings by about one power of $\alpha_s$.  The coefficients for improved staggered quarks with improved glue are shown in table \ref{T:Asqtad}; all are around 1, decreasing with $\zeta^2$, but the final column is zero at this order.  

\section{New Staggered Quark Actions}
This calculation gives us a tool to perturbatively investigate the taste-changing in tree-level actions. We investigated several significant modifications to the improved staggered quark action to try and reduce those coefficients and thereby avoid some or all of the four-quark operators. 
We tried partial re-unitarisation, adding an irrelevant term, and smearing the Lepage term from $\frac{\Delta_\rho^2}{4}$ to $(1\!+\!a\Delta^{+}_\rho\!-\!a\Delta^{-}_\rho)\frac{\Delta_\rho^2}4$.
The best modification was this Broadened Lepage Term [BLT] which reduces to negligible the more expensive ``3-Link'' contact terms as shown in the lower half of table~\ref{T:BLT}. Some early simulation results for pions splittings with this action are already available~\cite{Eduardo}.  

To go further and remove the remaining one-loop taste-changing terms requires implementing the contact terms in a simulation.
\begin{table}
\begin{center}
{\small
\begin{tabular}{ccccc}
\hline
&\multicolumn{4}{c}{1-Link}\\
$\zeta^2$&\multicolumn{2}{c}{Octet Colour}&\multicolumn{2}{c}{Singlet Colour}\\ 
\hline
\raisebox{-1.5mm}[0pt][0pt]{1}& 0.880i  & 0.500i & 0.643i  &          \\[-1mm]
     &\textcolor{gray}{$5\mu\!\otimes\!5$}
               &\textcolor{gray}{$5\mu\nu\!\otimes\!5\nu$}
                                  &\textcolor{gray}{$5\mu\!\otimes\!5$} & \\[0.5mm]
\raisebox{-1.5mm}[0pt][0pt]{2}& 0.435i  & 0.438i & 0.217i  &          \\[-1mm]
     &\textcolor{gray}{$\nu\!\otimes\!\mu\nu$}
               &\textcolor{gray}{$5\nu\!\otimes\!5\mu\nu$}
                                  &\textcolor{gray}{$\nu\!\otimes\!\mu\nu$} & \\[0.5mm]
\raisebox{-1.5mm}[0pt][0pt]{3}& 0.335i  & 0.409i & 0.244i  &          \\[-1mm]
     &\textcolor{gray}{$\mu\nu\!\otimes\!\nu$}
               &\textcolor{gray}{$1\!\otimes\!\mu$}
                                  &\textcolor{gray}{$\mu\nu\!\otimes\!\nu$} & \\[0.5mm]
\raisebox{-1.5mm}[0pt][0pt]{4}& 0.300i&\raisebox{-1.5mm}[0pt][0pt]{--}&0.220i&\raisebox{-1.5mm}[0pt][0pt]{--}                                                                    \\[-1mm]
     &\textcolor{gray}{$5\mu\!\otimes\!5$}&&\textcolor{gray}{$5\mu\!\otimes\!5$}& \\[0.5mm]
 \hline
$\zeta^2$&\multicolumn{4}{c}{3-Link}\\
\hline
\raisebox{-1.5mm}[0pt][0pt]{1}& 0.404i  & 0.518i  & 0.295i  &           \\[-1mm]
    &\textcolor{gray}{$1\!\otimes\!5\mu$}
              &\textcolor{gray}{$\mu\nu\!\otimes\!5\nu$} 
                                  &\textcolor{gray}{$1\!\otimes\!5\mu$}& \\[0.5mm]
\raisebox{-1.5mm}[0pt][0pt]{2}& 0.228i  & 0.364i  & 0.166i  &           \\[-1mm]
    &\textcolor{gray}{$\nu\!\otimes\!5\mu\nu$}
              &\textcolor{gray}{$5\nu\!\otimes\!\mu\nu$}
                                  &\textcolor{gray}{$\nu\!\otimes\!5\mu\nu$}&\\[0.5mm]
\raisebox{-1.5mm}[0pt][0pt]{3}& 0.198i  & 0.198i  & 0.145i  &           \\[-1mm]
    &\textcolor{gray}{$5\mu\nu\!\otimes\!\nu$}
              &\textcolor{gray}{$5\!\otimes\!\mu$}
                                  &\textcolor{gray}{$5\mu\nu\!\otimes\!\nu$} & \\[0.5mm]
\raisebox{-1.5mm}[0pt][0pt]{4}&\raisebox{-1.5mm}[0pt][0pt]{--}& 0.190i  &\raisebox{-1.5mm}[0pt][0pt]{--}&
                                                                        \\[-1mm]
    && \textcolor{gray}{$\mu\!\otimes\!5$}&& \\[0.5mm]\hline
    &$C=1$&\multicolumn{1}{c}{$C=-1$}&$C=1$&$C=-1$\\
\end{tabular}}
\caption{\label{T:Asqtad} Coefficients of one-loop taste-changing counter-terms for the Improved Staggered action, Improved Glue.}
\end{center}
\end{table}
Each four-quark operator can be individually removed in a simulation with the following trick:
\begin{align}
 \Delta\mathcal{L}= ic J\phi + \frac12 \phi^2 \qquad \equiv \qquad\Delta\mathcal{L} = \frac12c^2 J^2
\end{align}
because the scalar field $\phi$ is non-propagating.  This addition requires (8+1)$_\mu$ $\phi$'s per lattice site, but $icJ\phi$ terms are all one-link currents like the action, and can be easily accumulated in the calculation of the improved action with no change to the inverter.  However the coefficients are imaginary and therefore may cause positivity problems because $icJ\phi$ will be Hermitian unlike the rest of the quark action, in which case simulation with real coefficients and without allows extrapolation in the amount of taste-changing.

\section{Summary}
Taste-changing interactions are now well understood and can be reduced.  We recommend the Improved Staggered BLT action as a better tree-level improvement.  One-loop corrections allow alteration of the amount of taste-breaking either to reduce it or to quantifiably estimate the errors it induces.  Comparison perturbative and simulation results with the extra one-loop terms vs the HYP action are expected soon.
\begin{table}
\begin{center}
{\small
\begin{tabular}{ccccc}\hline
&\multicolumn{4}{c}{1-Link}\\
$\zeta^2$&\multicolumn{2}{c}{Octet Colour}&\multicolumn{2}{c}{Singlet Colour}\\ 
\hline
  1  & 0.947i   & 0.525i  & 0.692i   &          \\
  2  & 0.433i   & 0.397i  & 0.316i   &          \\
  3  & 0.281i   & 0.249i  & 0.207i   &          \\
  4  & 0.146i   &   -     & 0.105i   &   -      \\
 \hline
$\zeta^2$&\multicolumn{4}{c}{3-Link}\\
 \hline
 1  & 0.140i   & 0.173i  & 0.102i  &           \\
 2  & 0.084i   & 0.083i  & 0.061i  &           \\
 3  & 0.061i   & 0.089i  & 0.044i  &           \\
 4  &   -      & 0.069i  &   -     &           \\\hline
\end{tabular}}
\end{center}
\caption{\label{T:BLT} Coefficients of one-loop taste-changing counter-terms for the Improved Staggered action with Broadened Lepage Term [BLT], Improved Glue.}
\end{table}

\subsection*{Acknowledgement}
QM and GPL are supported by the National Science Foundation under Grant No. PHY-0098631, HDT by the Natural Sciences and Engineering Council of Canada, and JH, CD and EF acknowledge European Community's Human potential programme under HPRN-CT-2000-00145 Hadrons/LatticeQCD.
\end{fmffile}


\begin{thebibliography}{9}
\bibitem{Blum:1996uf} T.~Blum {\it et al.}, 
Phys.\ Rev.\ D {\bf 55} (1997) 1133, [hep-lat/9609036].
\bibitem{Lepage:1998vj} G.~P.~Lepage, 
Phys.\ Rev.\ D {\bf 59} (1999) 074502, [hep-lat/9809157].
\bibitem{Toussaint:2001zc} D.~Toussaint, 
Nucl.\ Phys.\ Proc.\ Suppl.\  {\bf 106} (2002) 111, [hep-lat/0110010].
\bibitem{Hein:2001kw} J.~Hein, Q.~Mason, G.~P.~Lepage and H.~Trottier, 
Nucl.\ Phys.\ Proc.\ Suppl.\  {\bf 106} (2002) 236, [hep-lat/0110045],
also W. Lee and S. Sharpe in these proceedings, [hep-lat/0208036].
\bibitem{Anna} A.~Hasenfratz, plenary in these proceedings.
\bibitem{Lee:1999zx}
W.~J.~Lee and S.~R.~Sharpe, 
Phys.\ Rev.\ D {\bf 60} (1999) 114503, [hep-lat/9905023].
\bibitem{Bernard:2001yj} C.~Bernard  [MILC Collaboration], 
Phys.\ Rev.\ D {\bf 65} (2002) 054031, [hep-lat/0111051],
updated in these proceedings [hep-lat/0209066].
\bibitem{Eduardo} E.~Follana {\it et al}, these proceedings, [hep-lat/0209122].
\end{thebibliography}
\end{document}